\documentclass[reprint,preprintnumbers,nofootinbib,amsmath,amssymb,floatfix,prl]{revtex4-2}

\usepackage{float}
\usepackage{graphicx}
\usepackage{bm}
\usepackage{hyperref}
\usepackage{xspace}
\usepackage[capitalise]{cleveref}
\usepackage{lmodern}
\usepackage{microtype}
\usepackage[T1]{fontenc}
\usepackage{siunitx}
\usepackage{subfig}
\usepackage{booktabs}
\usepackage[algoruled,lined,linesnumbered,longend]{algorithm2e}

\hypersetup{
    colorlinks=true,
    allcolors=[rgb]{0.26,0.41,0.88},
}

\SetKwFor{KwIn}{initialize}{}{}{}

\newcommand{\Z}{\ensuremath{\mathcal{Z}}\xspace}

\newcommand{\Pg}[2]{P\mathopen{}\left(#1\,\rvert\, #2\right)\mathclose{}}

\newcommand{\anyg}[3]{\ensuremath{#1\mathopen{}\left(#2\,\rvert\, #3\right)\mathclose{}}\xspace}

\usepackage[usenames]{xcolor}

\begin{document}

\title{Hunting for bumps in the margins}

\author{David Yallup}
\email{dy297@cam.ac.uk}
\affiliation{Cavendish Laboratory \& Kavli Institute for Cosmology, University of Cambridge,\\ JJ Thomson Avenue, Cambridge, CB3~0HE, United Kingdom}
\author{Will Handley}
\affiliation{Cavendish Laboratory \& Kavli Institute for Cosmology, University of Cambridge,\\ JJ Thomson Avenue, Cambridge, CB3~0HE, United Kingdom}


\begin{abstract}
Data driven modelling is vital to many analyses at collider experiments, however the derived inference of physical properties becomes subject to details of the model fitting procedure. This work brings a principled Bayesian picture -- based on the marginal likelihood -- of both data modelling and signal extraction to a common collider physics scenario. First the marginal likelihood based method is used to propose a more principled construction of the background process, systematically exploring a variety of candidate shapes. Second the picture is extended to propose the marginal likelihood as a useful tool for anomaly detection challenges in particle physics. This proposal offers insight into both precise background model determination and demonstrates a flexible method to extend signal determination beyond a simple bump hunt.
\end{abstract}

\maketitle

\section{Introduction}\label{sec:intro}
The searches and measurements taking place at the energy frontiers of collider physics take place in a data rich environment. Theoretical calculations of backgrounds in these analyses are compromised by the complexity of the full theory compared to what is computationally feasible. The problem of modelling these background processes is often further complicated by experimental systematics that are accounted for in-situ from collider data~\cite{ATLAS:2022swp}. These factors motivate usage of parametric models, fit to data, to perform the background model estimation~\cite{ATLAS:2019fgd,CMS:2019gwf,CMS:2021kom,ATLAS:2020pvn}. As a result, the search for (or measurement of) rarer signal processes amongst the background becomes sensitive to this data driven fit. There will be two main sources of uncertainty on the background model in such an analysis; parametric and modelling uncertainties. Parametric uncertainty arises due to the uncertainty on the input data points that the model is fit to, leading to variance of the parameters of a given model. Modelling uncertainty arises due to the potential for model misspecification biasing the resulting inferred functional form, this is generally much harder to estimate. Typically this is achieved by independently fitting multiple models and comparing the results of the individual fits, however the resulting recipe to select the model can become rather ad-hoc. Without quantifiable justification, the uncertainty from the model selection is difficult to precisely represent.

Bayesian inference offers a natural framework to discuss both modelling and parametric uncertainties~\cite{Trotta:2008qt}. Parameter uncertainty is naturally expressed in the posterior distribution of the parameters of interest. Model uncertainty is captured in the marginal likelihood, or \emph{evidence}, of a Bayesian calculation. Comparison of evidences between models gives a concrete method of Bayesian model selection, and can be used to define Bayesian model averages which effectively ensemble the modelling uncertainty. Alternatively (but giving equivalent results) the posterior can be extended over hyperparameters governing the choice of the model itself -- for example the family of functions composing the model or the number of basis functions. Such hyperparameters can be thought of as being discrete in nature, and understanding how to deal with discrete hyperparameters is one of the core goals of this study.

Numerically marginalising over discrete parameters is a computationally difficult task as discrete model choices induce discontinuous and multimodal likelihood surfaces. Nested Sampling (NS)~\cite{skilling} is a well established method to perform numerical marginalisation over such a target likelihood function. Bayesian sparse reconstruction (BSR)~\cite{bsr} has been demonstrated to perform the marginalisation over mixtures of discrete and continuous model parameters and hyperparameters in the context of astronomical imaging.

In this analysis the framework of BSR is applied to a toy problem representing a typical case of modelling a smoothly falling background process. The resulting composition of the model is examined and contrasted with existing maximum likelihood based approaches, motivating the utility of a Bayesian model in this context. To further highlight the unique opportunities afforded by NS in tackling multimodal sampling problems, a search for generic resonances -- a bump hunt -- is considered on top of the discrete marginalised background model. Signal analysis in particle physics has been cast as a Bayesian model comparison problem in plenty of previous work (see \cite{Fowlie:2019ydo} and references therein), by extending these kinds of signal analyses over a discrete marginalised background model, a method with potential to tackle more subtle shape discrimination is proposed. The result is an analysis pipeline defining a powerful, fully Bayesian method for anomaly detection with well-calibrated background modelling uncertainties. 

\section{Models for binned data}

Measuring an observable that is dominated by a single background process that has a smoothly falling rate as a function of the observable is a common scenario in many LHC physics analyses. The searches for resonances in dijet mass spectra at the ATLAS and CMS detectors~\cite{ATLAS:2019fgd,CMS:2019gwf} are flagship searches at the LHC that center on modelling the smoothly falling QCD background process. These analyses are particularly sensitive to the issue of extrapolating the fitted model into the poorly constrained tail of the spectrum. Precision measurements of Higgs properties in the diphoton decay channel with the ATLAS and CMS detectors~\cite{CMS:2021kom,ATLAS:2020pvn} provide a complementary example where the result hinges on the precision of the background model in a window as a function of the observable, surrounded and constrained by high statistics measurements of the diphoton mass spectrum. This study blends features from both these sources, examining the precise construction of a background model for a Higgs diphoton style measurement, then using this to perform an anomaly detection task reminiscent of a search analysis. In the remainder of this section the likelihood for this type of observable is defined, and the relevant basis functions to build the background model are defined.


\subsection{Likelihoods from histograms}

A single measurement channel consists of $M$ histogram bins in a measured property of collision events, $x$ (the dijet or diphoton mass for example). With bins indexed by $i$, data is collected in the $i$-th bin by counting events occurring in a range from the lower bin edge to upper bin edge. For simplicity the bin can instead be characterized by its mid-point, $x_i$, and the degree of freedom associated with the bin width can be effectively integrated out. Each bin records an observed event count, $n_i$, and the background model predicts this count as a function of the measured property, $b(x_i)$. The likelihood for the set of data observations $n_i$, given a background model parameterized by $\Theta$ can be taken as a product of independent Poisson distributions, 
\begin{align}\label{eq:like}
    \anyg{P}{n_i}{\Theta} = \prod_i^M  b(x_i,\Theta)^{n_i} \frac{e^{-b(x_i,\Theta)}}{n_i!}\,.
\end{align}
Additional constraint terms, nuisance parameters or parameters of interest on signal models can then be incorporated. This work initially explores the problem of inferring a background-only model in \cref{sec:datadrivenhiggs} to establish the challenges of surveying discrete parameters in models. In \cref{sec:bumphunt} this likelihood is then extended to consider a signal model in the guise of an anomaly detection task for generic signal models. 

The smoothly falling backgrounds in this problem can be modelled as a sum of simple basis functions $\phi$,
\begin{align}
    b(x_i,\Theta) = b(x_i,\theta,N,\phi) = \sum^N_k \phi_k(x_i,\theta)\,.
\end{align}
The background model is dependent on three factors in addition to the input $x_i$, redefining the background model parameters as $\Theta=\{ \theta,N,\phi\}$. These three components can be identified as -- the number of basis functions $N$, the parameters defining the basis functions $\theta$ and the family of basis functions $\phi$. The parameters $N$ and $\phi$ are identified as the discrete hyperparameters of the background model, and
the inference can either be extended over these parameters or they can be specified and conditioned on. A frequentist approach to deal with inference extended over nuisance parameters is to \textit{profile} them~\cite{Dauncey:2014xga}. Profiling a parameter removes the associated degree of freedom by conditioning on the maximum likelihood estimator for the parameter. Profiling discrete parameters is challenging, achieved by considering a model ensemble conditioned on the set of discrete model choices, which can be used to build an envelope of profiled curves over the continuous parameters. A method for constructing this ensemble is demonstrated in \cref{sec:mlmethod}.

A Bayesian interpretation of this problem instead marginalises over these discrete hyperparameters, inverting the likelihood using Bayes theorem to get a posterior distribution for the model parameters, $\Pg{\theta,N,\phi}{n_i}$. By marginalising over all parameters and hyperparameters in the model, a background model can be constructed that removes dependence on these choices by sampling directly from this full posterior distribution. An example of this marginalisation is demonstrated in \cref{sec:samplingmethod}, emphasising the role of the evidence as an alternative tool for building model averages. 

Methods involving Bayesian non-parametric models have received increased attention recently~\cite{Frate:2017mai}, giving strong motivation to consider the contrasting adaptive basis function method presented in this work. Much of the attraction of Gaussian Processes derives from the fact that the expression for the marginal likelihood of such models is directly calculable as a function of the training data from simple linear operations. A marginal likelihood calculated for a set of adaptive basis functions -- as presented in this work -- retains the strengths of promoting functional diversity and embodying automatic relevance determination encapsulated by the marginal likelihood, whilst negating conceptual issues arising from using non-parametric methods in this context. When used for anomaly detection tasks, non-parametric models require the length scale over which the anomaly exhibits to be well separated from the length scale of the background process. Whilst this is generally achievable for the types of narrow width signals considered in this demonstration, posing the modelling problem as a Bayes ratio enables the data driven background models to be extended to less well separated model comparison tasks.

\subsection{Smoothly falling functions}\label{sec:prior}

\begin{figure}
    \includegraphics[]{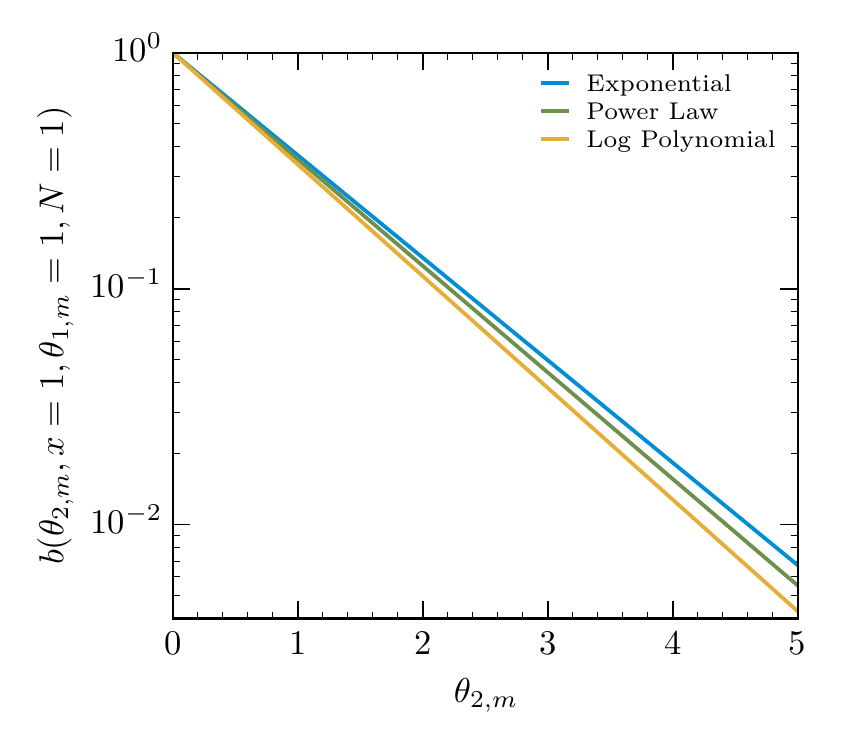}
    \caption{Value of background model at $b(\theta_{2,m},x=1,\theta_{1,m}=1,N=1)$ as a function of the exponent parameter $\theta_{2,m}$ for each discrete choice of $\phi$.\label{fig:ymax}}
\end{figure}

In order to construct a suitable set of basis functions to examine, a weak prior belief that the background model should be both constantly decreasing and smooth is initially considered. Following similar analyses conducted at collider experiments, three distinct candidate families of basis function are considered; Exponential (exp), Power law (power), and Log Polynomial (logpoly) functions. A more complete set of such functions has been demonstrated for similar physical use cases in the context of radio astronomy~\cite{Bevins:2020jqf}, and applied to similar bump hunting style problems in the same field~\cite{deLeraAcedo:2022kiu}. The three function families chosen are by no means an exhaustive list, being an arbitrary choice inspired by a priori reasonable candidates, leaving the task of selecting and weighting these candidate models to the data. These three choices of function family can be defined in terms of the remaining background model parameters as,
\begin{align}\label{eq:bases}
    \phi= \mathrm{exp}: \quad & b(x,\theta, N) = \sum_{m=1}^{N_{\mathrm{max}}} \theta_{1,m} \exp (-\theta_{2,m} x)\,, \\
    \phi= \mathrm{power}: \quad & b(x,\theta, N) = \sum_{m=1}^{N_{\mathrm{max}}} \theta_{1,m}  (x+1)^{-1.5\theta_{2,m}}\,, \\
    \phi= \mathrm{logpoly}: \quad & b(x,\theta, N) = \sum_{m=1}^{N_{\mathrm{max}}} \theta_{1,m} \log(x+e)^{-4\theta_{2,m}}\,.
\end{align}
The data is transformed such that $x_i\in[0,1]$ and $n_i$ scaled such that it has unit variance. The factors and translations appearing in these function definitions are required to ensure that the three families to admit similarly expressive functions. The $x$ axes of all functions are shifted such that ${b(\theta_{2,m},x=0,\theta_{1,m}=1,N=1)=1}$, and the exponents scaled such that ${b(\theta_{2,m},x=1,\theta_{1,m}=1,N=1)}$ has a similar range for all families. This similarity in range is illustrated in Figure~\ref{fig:ymax}. The \emph{whitening} of the data ensures that, given the chosen prior ranges on the functions, only sensible realisations of the background function are considered. When evaluating the likelihood, the functions are transformed back to the observed counts.

Each basis function in \cref{eq:bases} is characterized by two parameters; an amplitude parameter $\theta_{1,m}$ and an exponent parameter $\theta_{2,m}$. The exponent parameter is given a uniform prior on $[0,5]$. The amplitude parameter is given a \emph{sorted} (more formally, a forced identifiability prior~\cite{PhysRevD.100.084041}) uniform prior on $[-10,10]$. Allowing negative amplitudes relaxes the smooth constantly decreasing criteria in order to match the maximum likelihood implementation, however this is trivial to reinstate for a model with stronger inductive bias (as done in \cref{sec:bumphunt}).
The discrete hyperparameter $N$ is given a uniform prior over integers $[1,N_\mathrm{max}]$, with $\phi$ being effectively given a uniform prior over integers $[1,3]$ indexing the family choice. The resulting sampling space is of dimension $(N_\mathrm{max}\times 2 + 2)$, with previous applications of BSR finding it sufficient to mask inactive basis parameters when $N<N_\mathrm{max}$, and use the choice of $\phi$ as the key for a dictionary learning problem. Using a uniform prior on $N$ removes the sparsity promotion central to BSR, this can be easily reinstated by using an exponential prior on $N$ for example.

\section{Data driven Higgs background models}\label{sec:datadrivenhiggs}

A toy dataset illustrative of a diphoton background modelling problem can be constructed using event generators. The leading order diphoton \texttt{MEGammaGamma} built-in matrix element generator in Herwig 7~\cite{Bahr:2008pv,Bellm:2015jjp}, is used to generate particle level LHC events. Rivet~\cite{Bierlich:2019rhm} is used to analyse the events and produce a target histogram of the diphoton mass, $m_{\gamma\gamma}$, in 80 bins uniformly spaced on $m_{\gamma\gamma}\in[100,180]$ GeV. The events are simulated at $\sqrt{s}=13$~TeV, and projected to a dataset of size integrated luminosity, $\mathcal{L}=10~\mathrm{fb}^{-1}$. The generator is run until the Monte Carlo error in each bin is negligibly small. In order to simulate an experimental set of observations, the simulated $n_i$ in each bin is taken as the mean of a Poisson from which a toy ``observed'' $n_i$ is drawn. The model inference is performed over parameters and hyperparameters of the candidate background models with ranges and priors as detailed in \cref{sec:prior}, with $N_\mathrm{max}$ being set to 3. This section demonstrates a comparison of a maximum likelihood based method as well as the proposed Bayesian marginalisation method. 

\subsection{Maximum likelihood based methods}\label{sec:mlmethod}
The standard toolkit to estimate both the point estimate of the model parameters, and the various uncertainties, is based on estimating the parameter values that maximise the likelihood given in \cref{eq:like}. Conditioning on a choice of discrete hyperparameters, the values of the continuous basis parameters that maximise the likelihood (subject to the constraints and scaling applied in \cref{sec:prior}) can be found by a variety of optimization methods such as gradient descent. The value of the maximum log-likelihood for each set of discrete parameter choices are listed in \cref{tab:results}. Whilst the likelihood is a useful tool to pick the continuous parameters, there is an obvious flaw in using it to pick the value of $N$ -- larger values of $N$ will always be preferred. A potential resolution that is often employed in maximum likelihood approaches to modelling uncertainties is to introduce a correction based on the dimensionality of the model to counterbalance this preference for overfit models.
There are a number of ways to motivate a penalty term, previous exploration of the correction term in particle physics applications has considering the magnitude of this correction as a tunable parameter~\cite{Dauncey:2014xga}. In this example the Akaike information criterion (AIC) is used as a benchmark, one of the candidate corrections considered in previous work. The AIC is derived from the maximum likelihood value, $\hat{\mathcal{L}}$, as,
\begin{equation}
    \label{eq:aic}
    -\mathrm{AIC} = 2 \ln {\hat{\mathcal{L}}} - 2k\,,
\end{equation}
where $k$ is the number of free parameters in the model (equal to $2N$ for all models considered). The discrete profiling method uses this idea to correct the profiled values of parameters of interest. In a simpler background only fit, it is illustrative to consider another quantity that can be derived from the AIC of each model, the relative likelihood,
\begin{equation}
    \label{eq:p_aic}
    \mathcal{L}(N,\phi)_{\mathrm{AIC}}=\exp \bigg( \frac{\mathrm{AIC}_{\mathrm{min}}- \mathrm{AIC}_{N,\phi}}{2} \bigg)\,,
\end{equation}
with the subscripts referring to the minimum AIC found across the entire range of $\{N,\phi\}$ and the specific choice in question. This is listed for the considered functions in \cref{tab:results}, giving an effective model composition based on the penalised maximum likelihood.


\setlength{\tabcolsep}{10pt}

\begin{table*}[t]
    \caption{Summary of comparison between marginalising and profiling based approaches for background composition. The probability is derived from the height and penalty as defined in \cref{eq:p_aic} for the profiling case and \cref{eq:marg_model} for the marginal case. Choices of $\{N,\phi \}$ contributing to the total model with a probability greater than 0.1 are listed in bold. }
    \label{tab:results}
    \smallskip
    \begin{center}
    \begin{small}
    \begin{tabular}{cccccccc}
      \toprule
      & & \multicolumn{3}{c}{Discrete Profiling} & \multicolumn{3}{c}{Discrete Marginalising} \\
      \cmidrule(lr){3-5} \cmidrule(lr){6-8} \\
      $N$ & $\phi$ & Height ($\ln\hat{\mathcal{L}}$)  & Penalty ($2k$)   & Probability & Height ($\langle \ln \mathcal{L} \rangle_P$)  & Penalty ($D_\mathrm{KL}$)  & Probability \\
      \midrule
      1 & exp & -366.03 & 4 & 0.01 &  -366.89 & 8.83 & 0.02 \\
      1 & power & -384.09 & 4 & $<0.01$ &  -385.09 & 8.9 & $<0.01$ \\
      1 & logpoly & -374.03 & 4 & $<0.01$ &  -374.98 & 8.84 & $<0.01$ \\
    \hline
      2 & exp & -359.99 & 8 &  \textbf{0.57} &  -361.53 & 11.18 &  \textbf{0.32} \\
      2 & power & -362.24 & 8 & 0.06 &  -362.99 & 10.99 & 0.09 \\
      2 & logpoly & -361.35 & 8 & \textbf{0.15} &  -362.29 & 12.38 & \textbf{0.15} \\

    \hline

      3 & exp & -360.09 & 12 &  0.07 &  -361.45 & 11.49 &  \textbf{0.25} \\
      3 & power & -360.11 & 12 & 0.07 &  -362.29 & 11.18 & \textbf{0.15} \\
      3 & logpoly & -360.08 & 12 & 0.07 &  -362.15 & 11.41 & \textbf{0.14} \\
      \bottomrule
    \end{tabular}
\end{small}
\end{center}
\end{table*}

\subsection{Sampling based methods}\label{sec:samplingmethod}
When uncertainty quantification on parameter values is required alongside extracting the best fit choices, the background model can be equivalently constructed by sampling from a typical likely set of parameters. This Bayesian interpretation of the background modelling problem is achieved by sampling across a defined range of parameter values, as given in \cref{sec:prior}, and using the density weighted by the likelihood to define the regions of parameter space of interest. The result is a \emph{discrete marginalised model}, and the sampling is performed by using the \textsc{PolyChord} implementation of Nested Sampling~\cite{Handley:2015vkr} to numerically marginalise over the multimodal likelihood. The resulting posterior for the discrete hyperparameters is listed in \cref{tab:results}, sampling from this distribution gives a weighted sample of background models which can be propagated through further inference machinery.

\begin{figure}
    \begin{center}
        \includegraphics[]{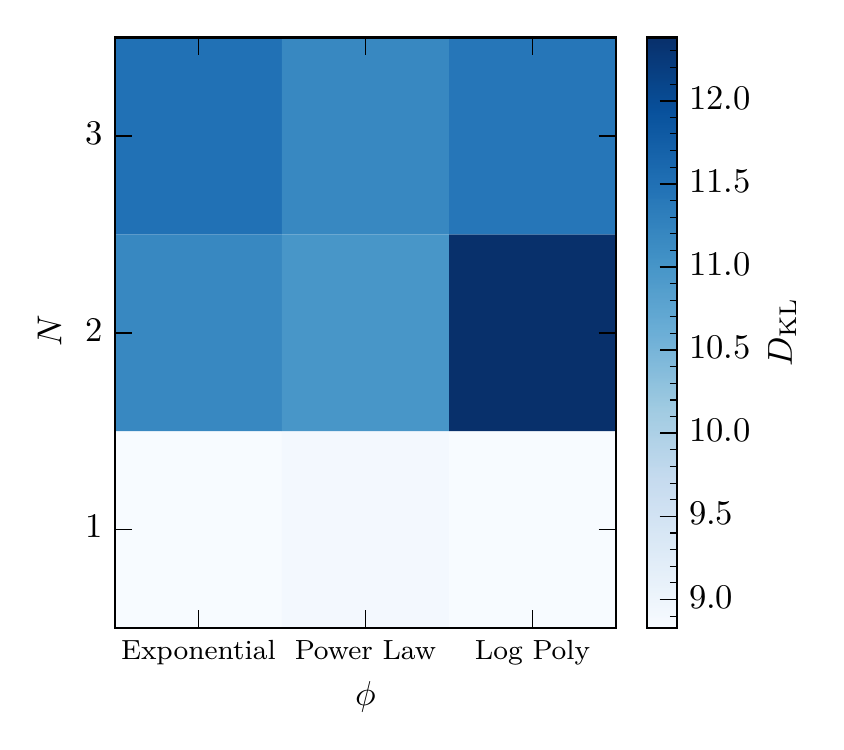}
    \end{center}
    \caption{Kullback-Leibler divergence, $D_\mathrm{KL}$, derived from the marginalised sample as a function of the discrete model parameters.\label{fig:dkl}}
\end{figure}

Using the same sample we can calculate the local evidence for each mode, noting that the problem could be equivalently formulated as a set of individual evidence calculations conditioned on each pair of discrete hyperparameter values. The local log-evidence of each mode can be calculated as,
\begin{equation}\label{eq:marg_model}
    \ln\Z  = \langle \ln \mathcal{L} \rangle_P - D_\mathrm{KL}\,,
\end{equation}
where the log-evidence, $\ln\Z$, is equal to the posterior weighted mean of the log-likelihood, $\langle \ln \mathcal{L} \rangle_P$, minus the Kullback-Leibler divergence, $D_\mathrm{KL}$, between the prior and the posterior~\cite{Hergt:2021qlh}. This equation can be viewed in analogy to the maximum likelihood construction given in \cref{eq:aic}; with the posterior mean likelihood playing the same role as the maximum likelihood in setting the ``height'' of each mode and the Kullback-Leibler divergence playing the role of the penalty term. The Kullback-Leibler divergence, $D_\mathrm{KL}$ is defined as the relative entropy between the prior, $\pi(\theta)$, and posterior, $P(\theta)$, as,
\begin{equation}
    D_\mathrm{KL} = \int d\theta P(\theta) \ln \frac{P(\theta)}{\pi(\theta)}\,.
\end{equation}
This quantifies the information gain between prior and posterior, and hence can be interpreted as an \emph{Occam penalty}~\cite{Hergt:2021qlh}, punishing models based on relative lack of information gain. The posterior weighted mean log-likelihood is listed in \cref{tab:results} and the corresponding Kullback-Leibler divergence calculated from the sample is shown in \cref{fig:dkl}. As the sample was extended over the hierarchical parameters, the posterior distribution listed in \cref{tab:results} can be estimated directly from the samples, which is what is listed. Alternatively, it is equivalent to instead sample each choice of $\{N,\phi\}$ as distinct models and then weight each model $i$ according to $\Z_i/\Z_\mathrm{sum}$. 

\subsection{A comparison of methods}
The previous sections derived two alternative methods to estimate a model composition for a diphoton background model. Inspection of the two model compositions in \cref{tab:results} show that both methods favour $N=2$ and $\phi=\mathrm{exp}$ functions, however the discrete marginalisation promotes a greater diversity in functional forms. The diversity of the total background model in terms of $\{N,\phi\}$ is dictated by the number of different entries in \cref{tab:results} with a significant probability, where $P(N,\phi)>0.1$ is used as a rough visual guide. The diversity along the $N$ axis of this table can be compensated for in the maximum likelihood method by tuning the penalty term used, the work of Dauncy et al.~\cite{Dauncey:2014xga} notes the difficulty in making general recommendations for such a tuning due to it's dependency on the studied application. It is equivalently possible to instead ``tune'' the sampling based method by introducing a sparsity promoting prior -- for example an exponential prior on $N$. There is also increased diversity in the model composition along the $\phi$ axis in the marginalising case, particularly visible for the functions with $N=3$. There is no clear mechanism to account for this effect in a discrete profiling framework. To summarize this comparison suggests that inference built from a maximum likelihood based method needs problem specific tuning and is challenged when faced with subtle shape differences.

\begin{figure}
    \includegraphics[]{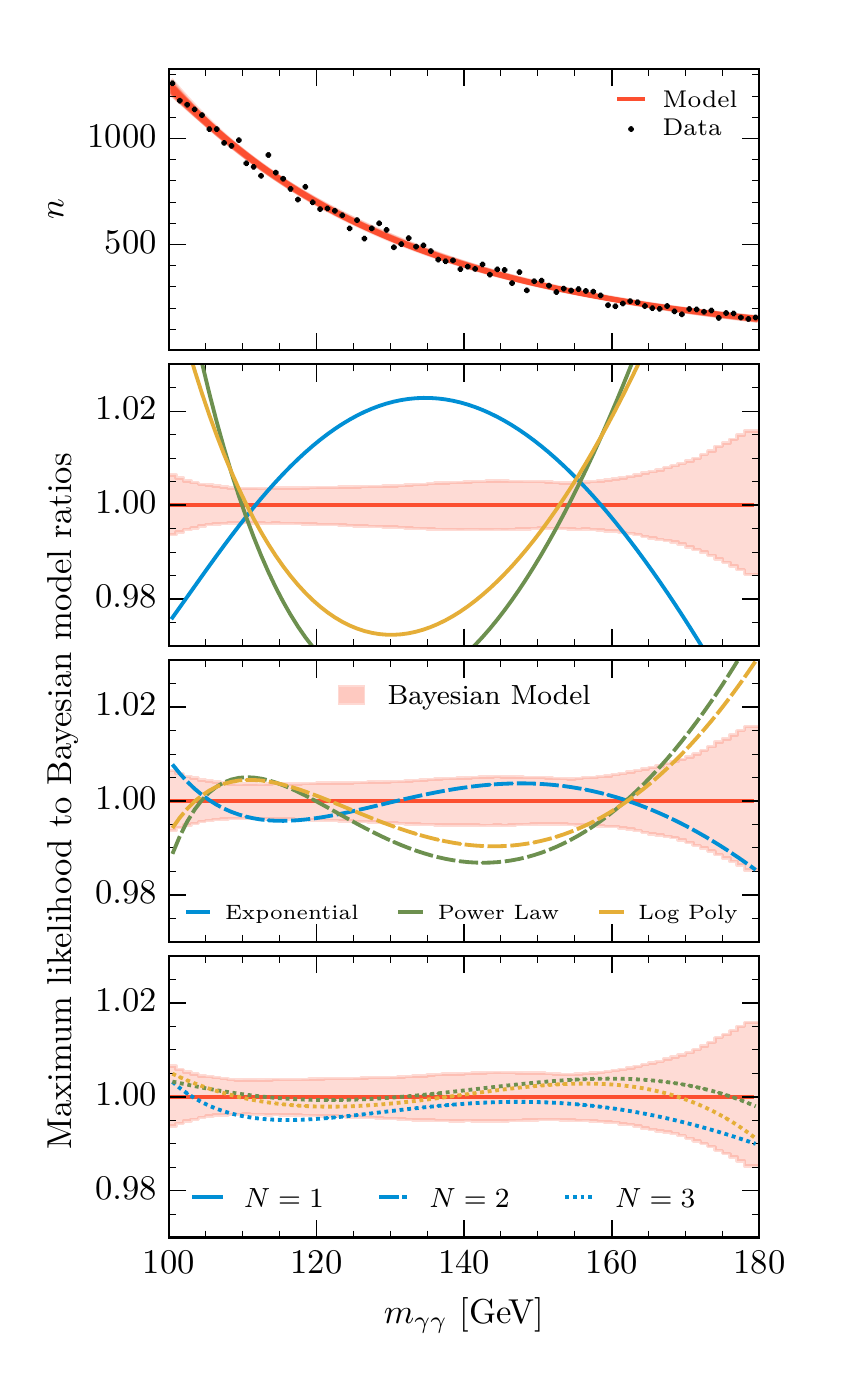}
    \caption{Prediction and data included in the inference problem for the background model in terms of the physical variables $\{n,m_{\gamma\gamma}\}$. The top panel displays the observed data and predicted spectrum resulting from 100 posterior samples of the Bayesian model in red. The three ratio panels show the ratio of maximum likelihood models to the mean prediction from the Bayesian ensemble for increasing values of $N$. The variance of the ensemble is used to predict a one standard deviation error band shown shaded in red.\label{fig:mggfit}}
\end{figure}

These two model compositions can be visually presented by examining how the inferred background models behave as a function of the physical observable, this is shown in \cref{fig:mggfit}. The Bayesian model prediction is the result of drawing 100 samples from the posterior and deriving an ensemble of model predictions from this set of parameter values -- this forms a posterior predictive distribution for the physical observable $m_{\gamma\gamma}$. The set of all 100 function predictions are shown in red in the top panel of the figure over the considered data. The posterior predictive distribution can be used to define mean model prediction and a one standard deviation uncertainty band on this. The bottom three ratio panels show the ratio taken to this Bayesian Model mean of both the full Bayesian error band, as well as the maximum likelihood solutions for each model choice.

Examining the composition of the background in detail is a useful validation of the method, however typically the composition of the background model is understood in terms of its influence on inferred parameters of a hypothesised signal model. It is conceptually simple enough to introduce an explicit parameterized signal model into \cref{eq:like} without introducing any particularly challenging computational overhead to this method, and this is explored in \cref{sec:bumphunt}.




\section{Bump hunting with marginal likelihoods}\label{sec:bumphunt}
Performing a Bayesian analysis of typical particle physics parameterized signal models is a well examined topic, with nuanced analysis needed to navigate issues such as the impact of choice of prior on physical parameters of interest~\cite{Fowlie:2019ydo}. The key quantity needed for a Bayesian analysis of a signal model is already derived, the evidence for a hypothesis as in \cref{eq:marg_model}. A generic parametric signal model can be introduced to the existing likelihood shown in \cref{eq:like} as,
\begin{align}\label{eq:like_signal}
    H_\psi: \quad  \anyg{P}{n_i}{\psi,\Theta} = \prod_i & \big(s(x_i,\psi) + b(x_i,\Theta)\big)^{n_i} \,\\
                                                    &  \cdot \frac{e^{-(s(x_i,\psi) +  b(x_i,\Theta))}}{n_i!}\,,
\end{align}
where the signal model, $s(x_i,\psi)$, is a function of the input $x$ data as well as a new set of parameter(s) of interest, $\psi$. Conditioning on a choice of $\psi$ specifies the hypothesis in question, $H_\psi$, with the likelihood in \cref{eq:like} being identified as the null hypothesis, $H_0$. 

Rather than further study the properties of derived intervals on signal models with strong priors -- or validating the properties of frequentist vs. Bayesian discovery metrics -- a generic anomaly detection task is considered. This scenario employs weak priors on signal parameters, and considers a likelihood with only a weak signal injected. This is a deliberately challenging scenario and is reflective of recent activity in the field trying to push the envelope of anomaly detection. By injecting a signal at around the noise threshold, a signal model that is multimodal in its parameter space is induced. This plays to a previously highlighted strength of a Nested Sampling based approach, the multimodal background model built earlier is still employed and a multimodal signal model on top of this creates a sampling problem that many common numerical sampling tools would struggle with. 

For the purposes of this study a generic Gaussian signal model is considered, defining the set of $\psi=\{A,\mu,\sigma\}$, as the amplitude, mean and variance of a Gaussian distribution. In this analysis all of these parameters are included into the marginalisation process, seeking to obtain the evidence for the hypothesis $\Z(H_\psi)=\Z_\psi$. In particular the search for a generic resonance signal across the entire spectrum is sought, commonly recognised as a bump hunt in particle physics~\cite{Choudalakis:2011qn}. In the following sections the same resonance search is applied in a null dataset (\emph{i.e.}, the data used in \cref{sec:datadrivenhiggs}) and a dataset with a true signal injected. The priors on the background model parameters are the same as described in \cref{sec:prior}, apart from the amplitude prior which is given a sorted uniform prior on $[0,5]$.


\subsection{Uncovering true positives}
\begin{figure}
    \includegraphics[]{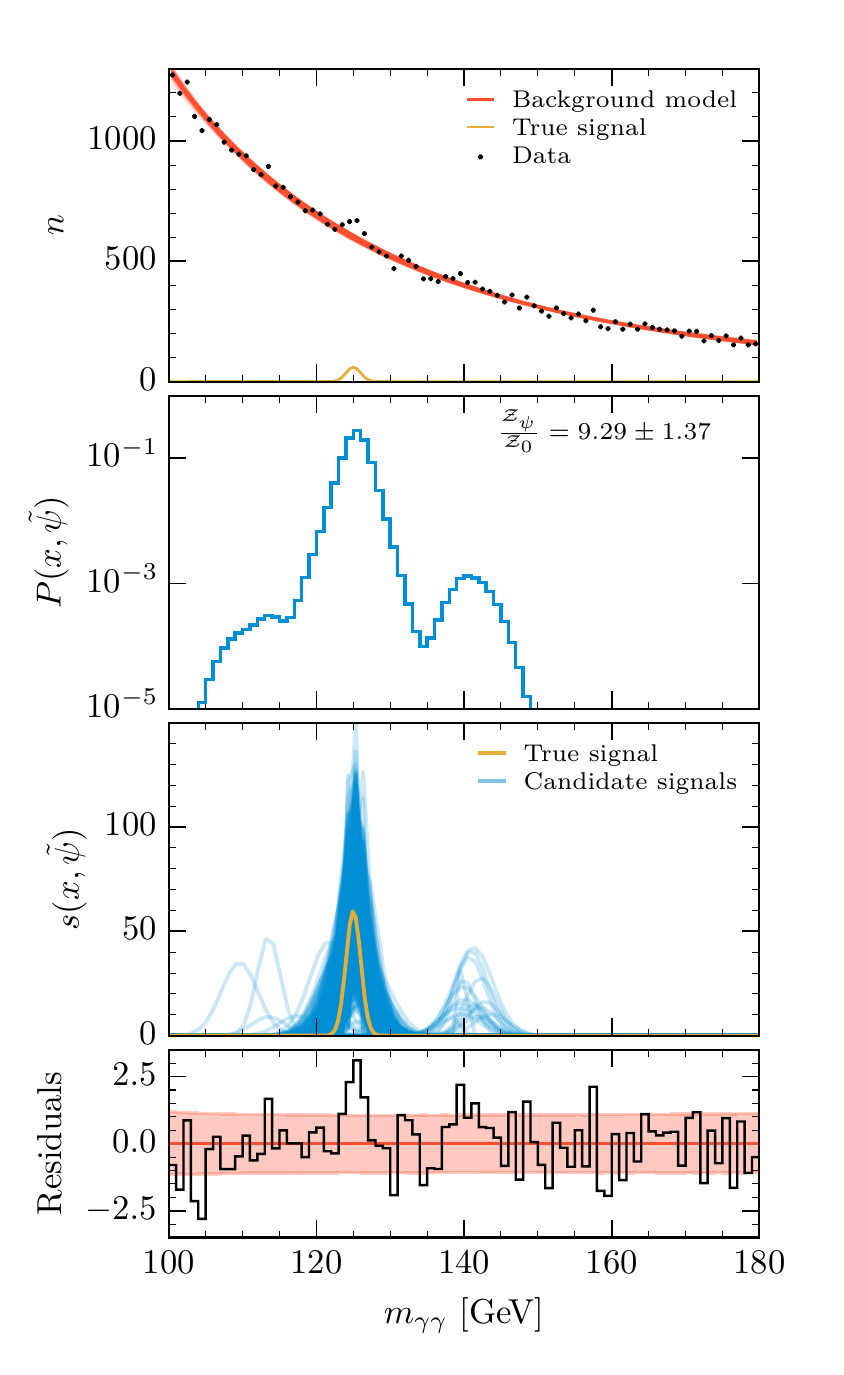}
    \caption{Summary of the sampled signal hypothesis, $H_\psi$ presented as a function of the physical observable, $m_{\gamma\gamma}$. This is a test to recover true positives, with a true signal model inserted to the background data.\label{fig:psi_predict}}
\end{figure}

First a check is performed to see if, upon injecting a ``true'' signal to the null dataset, a correct inference of the signal can be recovered. To perform both of these searches the evidence for the signal model, $\Z_\psi$, and the evidence for the null model, $\Z_0$, are calculated. To allow a generic search across the spectrum the following uniform priors are put on the components of $\psi$, 
\begin{itemize}
    \item $A$ - the signal amplitude is given a prior in a range $[0,500]$, noting that as the signal model has by construction an integral of 1, this corresponds to a signal model with cross section $[0,0.05]$ fb. The true injected value is $A=150$.
    \item $\sigma$ - the signal variance is given a prior in a range of $[0.5,3]$. With a unit bin width this restricts the search to a narrow resonance search, confining the signal to being contained mostly within $[1,6]$ bins. The true injected value is $\sigma=1.0$.
    \item $\mu$ - the signal location is given a prior range covering [$100,180$] GeV. This styles this search as a bump hunt across the full spectrum of data. The true injected value is $\mu=125$.
\end{itemize}
A summary of the results of calculating the marginal likelihood for signal and null hypotheses is shown in \cref{fig:psi_predict}. The four panels in this figure show (in descending order); the physical spectrum and background model (as previously displayed in the top panel of \cref{fig:mggfit}), the posterior predictive distribution for the physical observable under $H_\psi$, a set of 1000 candidate forms for the signal model itself sampled from the posterior, and lastly the data residuals normalised by the error. The key quantity that amortises the hunt for the signal hypothesis is the ratio of the evidences, $Z_\psi/Z_0=9.29 \pm 1.37$. This can be interpreted as ``betting odds''~\cite{Handley:2019tkm}, implying the data favouring the signal hypothesis over the null with an odds ratio of over 9 to 1. This global hypothesis odds is complemented by the calculated per-bin posterior signal model probability. By sampling a set of posterior samples for the signal model parameters, $\tilde{\psi}$, a set of candidate signal models can be constructed, $s(x,\tilde{\psi})$. The binned weighted sum of these posterior predictive distributions for $s$ gives the per-bin signal predictive posterior under $H_\psi$. In this example there is a clear preference for candidates around the true injected signal, however another potential mode for the signal model is weakly identified. The bottom panel of \cref{fig:psi_predict} shows the residuals -- the ratio of the data to the mean background model normalised to the variance of the background model. It is notable that the signal injected at $m_{\gamma\gamma}=125$ GeV consists of multiple adjacent bins where the data deviates from the model by over two standard deviations. The Bayesian significance implied by the ratio of $Z_\psi/Z_0$ is noticeably lower than one would naively expect, and the tension between this and the equivalent frequentist global significance has been described as the Bayes effect~\cite{Fowlie:2019ydo}. This tension is a significant open problem that merits further exploration building on both the NS application and discrete marginalising presented in this work.

\begin{figure}
    \includegraphics[]{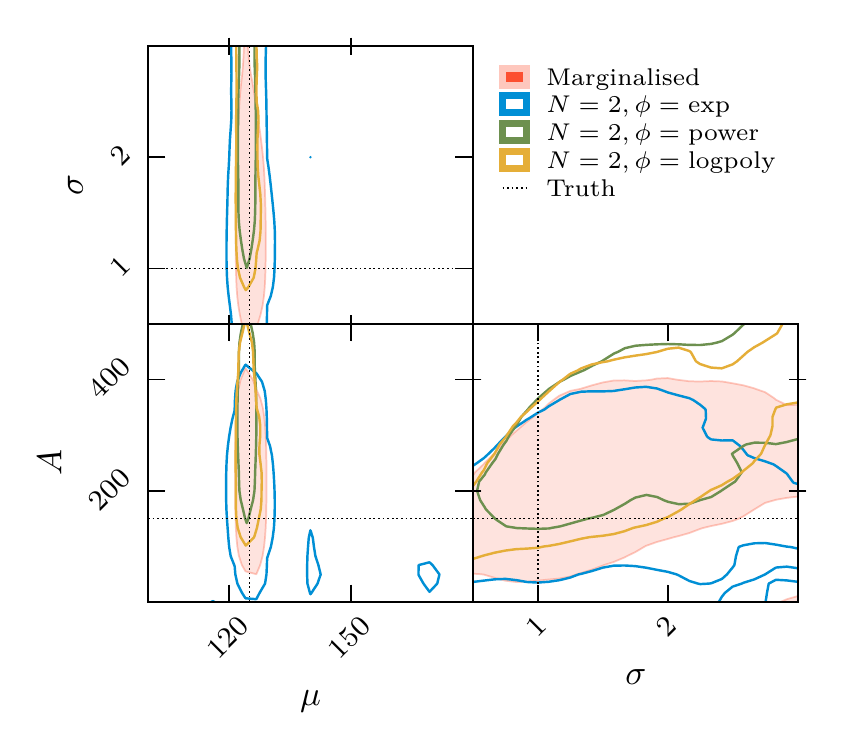}
\caption{Posterior distribution of signal parameters under $H_\psi$ showing the 95\% credibility regions for 3 fixed hyperparameter value background models as outlines, alongside a filled contour representing the corresponding region where marginalisation over the background hyperparameters takes place. The true values of the inserted signal model parameters are shown as dotted lines.\label{fig:signal_post}}
\end{figure}

Examining the model prediction as a function of the physical observables is a useful diagnostic, but the most natural language to discuss signal models is to calculate the posterior distributions for the parameters of the model. The corner plot for $\psi$ is shown in \cref{fig:signal_post}. In order to highlight the utility of the marginal background model as described in \cref{sec:samplingmethod}, this corner plot shows the 95\% credibility region posterior for the signal parameters under four background hypotheses. The full discrete marginalised background model is shown as a shaded contour, and comparison is drawn to three background models with fixed hyperparameters, corresponding to each of the three families considered and $N=2$. The dotted black lines show the true inserted signal parameter values. Of the three fixed background models, $\phi=\mathrm{exp}$ covers the true values well, but offers a worse resolution of the location and amplitude parameters. Both $\phi=\mathrm{power}$ and logpoly offer better resolution of the signal parameters but are close to failing to cover the true values. Marginalising over these models (and the other choices of $N$) results in a combined model that is able to retain the desirable features of the component backgrounds. In all cases the width of the signal model is poorly constrained but this conclusion is a feature of the toy problem, wherein the signal was injected at a level only slightly larger than Poisson noise. Conclusions on the various choices of fixed background hyperparameters is largely dependent on the generated pseudo-data used, but the overall method of marginalising out these choices is robust. Further investigation preferably on real data and incorporating more exhaustive comparison with results obtained via discrete profiling would be well motivated.


This example establishes a marginal likelihood calculation as a successful strategy to search for, and infer parameters of, anomalous bumps in smoothly falling spectra. More
in-depth study is needed to explore the threshold of signal magnitude (and variety of signal shapes) to establish
the false negative error rate for a given dataset.  Examining how precisely shape deviations can be extracted is pushed to new limits with the well calibrated in-situ data driven backgrounds composed in this work, opening up an attractive prospect for an unsupervised learning algorithm built on the marginal likelihood. This can be further extended by considering signal models with shapes that go beyond a simple Gaussian~\cite{Alvarez:2022kfr}, combining the power of the functionally diverse data driven background models exhibited in this work with more exotic signal shapes.

Pushing beyond the bump hunting paradigm is a topic of renewed interest more generally in the field~\cite{Kasieczka:2021xcg}, where signal models are pushed to undetectable levels when relying on simple bump hunting paradigms. Unsupervised machine learning methods generally attempt to overcome this by extending inference over a higher dimensional input data space, however in many cases these approaches rely on some variant of bump hunting in the latent space of a larger ML architecture. Adapting this work to such a scenario is an area with strong potential.

\subsection{Checking for false positives}
\begin{figure}
    \includegraphics[]{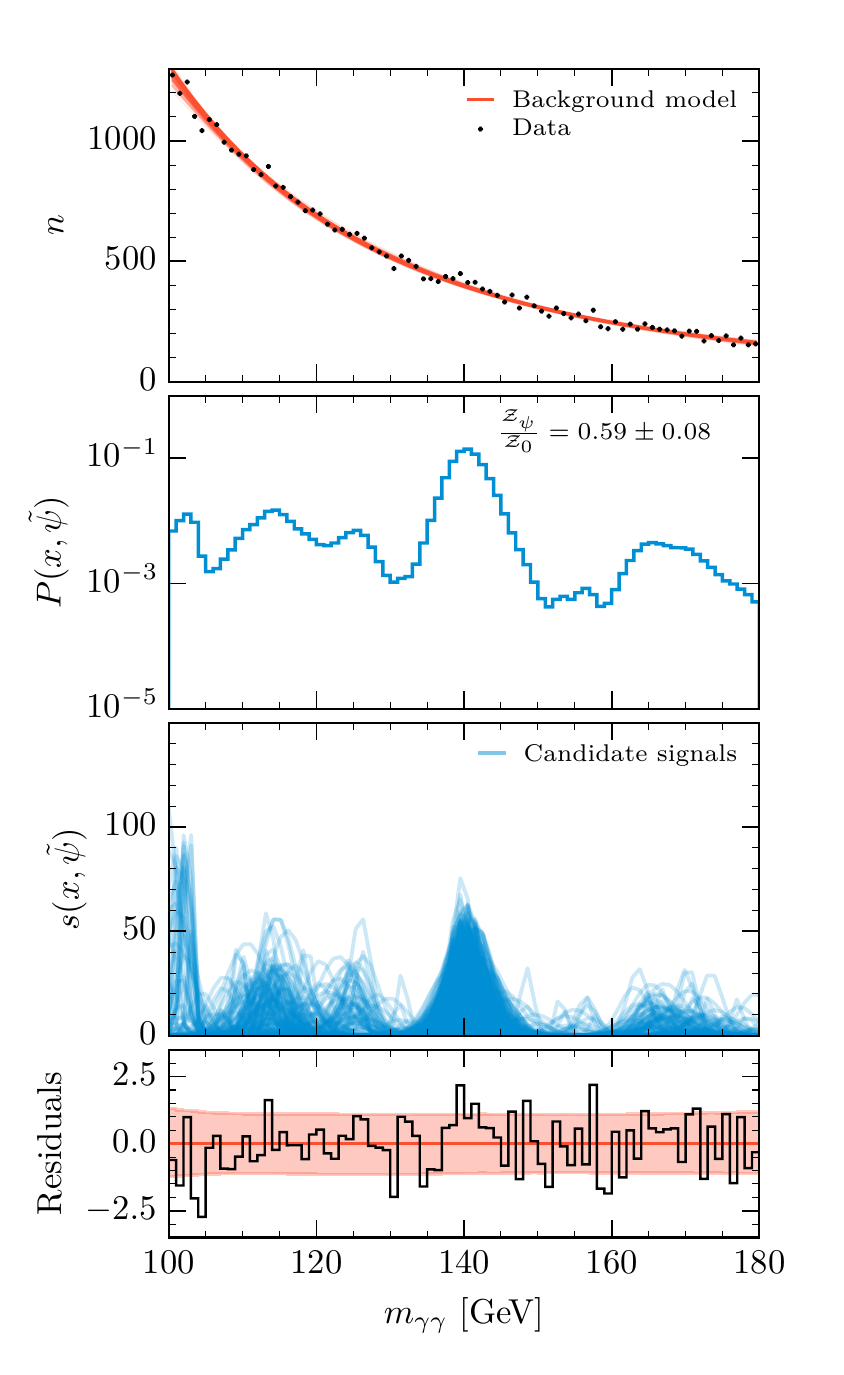}
    \caption{Summary of signal hypothesis as in \cref{fig:psi_predict}, in the regime where no signal has been injected.\label{fig:psi_predict_null}}
\end{figure}

Alongside exploring the ability to retrieve true positive injected signals, it is important to balance this with a consideration of the robustness in the face of false positive signal identification. This sort of false positive can be examined by running the proposed method on a spectrum with no signal injected. The summary of running a marginalisation of $H_\psi$, in the case where there is no signal injected, is shown in \cref{fig:psi_predict_null}, mirroring the structure of \cref{fig:psi_predict}. It is in examining this scenario that the strength of a Bayesian anomaly detection really shines. Whilst there are locally significant deviations, the overall evidence for the signal hypothesis disfavoured over the null -- by a probability ratio of $Z_\psi/Z_0=0.59 \pm 0.08$. The fully Bayesian method proposed in this work demonstrates the advantages of the Bayesian Occams razor end-to-end. In \cref{sec:samplingmethod} it was noted that Bayesian Occams razor optimally navigated background model complexity, punishing overly complex models based on relative lack of information gain. The principles of Occams razor are once again on display when the marginalisation is extended over signal models, allowing discrimination of signal models based on how economically they improve the description of the data. It is precisely this amortisation of models based on dimensional economy in comparison to the null, that render this a powerful yet well calibrated search method. 

\section{Conclusion}

Bayesian inference, with a particular focus on evidence calculations, gives a principled framework to handle comparison between models. By including hyperparameters governing choice of model directly into a Bayesian sampling framework we can simultaneously perform model comparison whilst sampling appropriate values of the background model parameters. The predictions can be treated simply as a single band covering all the necessary uncertainty on this background model and propagated to further inference machinery.

A toy example of inferring a data driven background model for a Higgs measurement was used to demonstrate the machinery in a familiar collider physics use case. The marginalised Bayesian model gives a principled uncertainty in terms of all the nuisance parameters -- including discrete hyperparameters -- of the model, capturing both modelling and parameter uncertainty simultaneously. Precision measurements and searches stand to benefit from the precise uncertainty and improved weighting of functionally diverse candidate models. 

The benefit for search analyses was demonstrated by extending the discrete marginalised background model to include a search for a generic anomaly using a Gaussian signal model. Calculation of evidence ratios in challenging scenarios reveals a well calibrated method for anomaly detection in particle physics spectra. This constructs a prototype analysis pipeline that can handle complex signal shapes alongside diverse data driven backgrounds. This pipeline presents a challenging sampling problem, but one that is handled well by the Nested Sampling tools employed in this analysis.

More broadly, evidence calculations and subsequent evidence weighted combinations of distinct models presents a statistically justifiable method to combine discrete model choices into a single unified model. Long standing comparisons in collider phenomenology, such as comparison of hadronization models or choice of hard process scale~\cite{Yallup:2022yxe}, can be viewed as a Bayesian model comparison using the framework presented in this work.

\section*{Acknowledgements}
    This work was supported by a Royal Society University Research Fellowship and enhancement award. The authors gratefully acknowledge input and comments from Mike Hobson on the construction of this manuscript.

    The \texttt{anesthetic} package~\cite{anesthetic} was used to appropriately manipulate the NS chain files to construct the figures throughout this work. The \texttt{fgivenx} package~\cite{fgivenx} was used to build the function error visualisations. A prototype of the analysis code is available at \href{https://github.com/yallup/discrete_marginal}{\texttt{[github.com/yallup/discrete\_marginal]}}

\bibliography{references}

\providecommand{\href}[2]{#2}\begingroup\raggedright\begin{thebibliography}{10}

\bibitem{ATLAS:2022swp}
{\scshape ATLAS} collaboration, \emph{{Tools for estimating fake/non-prompt
  lepton backgrounds with the ATLAS detector at the LHC}},
  \href{https://arxiv.org/abs/2211.16178}{{\ttfamily 2211.16178}}.

\bibitem{ATLAS:2019fgd}
{\scshape ATLAS} collaboration, \emph{{Search for new resonances in mass
  distributions of jet pairs using 139 fb$^{-1}$ of $pp$ collisions at
  $\sqrt{s}=13$ TeV with the ATLAS detector}},
  \href{https://doi.org/10.1007/JHEP03(2020)145}{\emph{JHEP} {\bfseries 03}
  (2020) 145} [\href{https://arxiv.org/abs/1910.08447}{{\ttfamily
  1910.08447}}].

\bibitem{CMS:2019gwf}
{\scshape CMS} collaboration, \emph{{Search for high mass dijet resonances with
  a new background prediction method in proton-proton collisions at $\sqrt{s}
  =$ 13 TeV}}, \href{https://doi.org/10.1007/JHEP05(2020)033}{\emph{JHEP}
  {\bfseries 05} (2020) 033}
  [\href{https://arxiv.org/abs/1911.03947}{{\ttfamily 1911.03947}}].

\bibitem{CMS:2021kom}
{\scshape CMS} collaboration, \emph{{Measurements of Higgs boson production
  cross sections and couplings in the diphoton decay channel at $
  \sqrt{\mathrm{s}} $ = 13 TeV}},
  \href{https://doi.org/10.1007/JHEP07(2021)027}{\emph{JHEP} {\bfseries 07}
  (2021) 027} [\href{https://arxiv.org/abs/2103.06956}{{\ttfamily
  2103.06956}}].

\bibitem{ATLAS:2020pvn}
{\scshape ATLAS} collaboration, \emph{{Measurement of the properties of Higgs
  boson production at $\sqrt{s}$=13 TeV in the $H\to \gamma\gamma$ channel
  using 139 fb$^{-1}$ of $pp$ collision data with the ATLAS experiment}},
  2020, \href{cds.cern.ch/record/2725727}{cds.cern.ch/record/2725727}.

\bibitem{Trotta:2008qt}
R.~Trotta, \emph{{Bayes in the sky: Bayesian inference and model selection in
  cosmology}}, \href{https://doi.org/10.1080/00107510802066753}{\emph{Contemp.
  Phys.} {\bfseries 49} (2008) 71}
  [\href{https://arxiv.org/abs/0803.4089}{{\ttfamily 0803.4089}}].

\bibitem{skilling}
J.~Skilling, \emph{{Nested sampling for general Bayesian computation}},
  \href{https://doi.org/10.1214/06-BA127}{\emph{Bayesian Analysis} {\bfseries
  1} (2006) 833 }.

\bibitem{bsr}
E.~Higson, W.~Handley, M.~Hobson and A.~Lasenby, \emph{Bayesian sparse
  reconstruction: a brute-force approach to astronomical imaging and machine
  learning}, \href{https://doi.org/10.1093/mnras/sty3307}{\emph{Monthly Notices
  of the Royal Astronomical Society} (2018) }
  [\href{https://arxiv.org/abs/1809.04598}{{\ttfamily 1809.04598}}].

\bibitem{Fowlie:2019ydo}
A.~Fowlie, \emph{{Bayesian and frequentist approaches to resonance searches}},
  \href{https://doi.org/10.1088/1748-0221/14/10/P10031}{\emph{JINST} {\bfseries
  14} (2019) P10031} [\href{https://arxiv.org/abs/1902.03243}{{\ttfamily
  1902.03243}}].

\bibitem{Dauncey:2014xga}
P.D.~Dauncey, M.~Kenzie, N.~Wardle and G.J.~Davies, \emph{{Handling
  uncertainties in background shapes}: {the discrete profiling method}},
  \href{https://doi.org/10.1088/1748-0221/10/04/P04015}{\emph{JINST} {\bfseries
  10} (2015) P04015} [\href{https://arxiv.org/abs/1408.6865}{{\ttfamily
  1408.6865}}].

\bibitem{Frate:2017mai}
M.~Frate, K.~Cranmer, S.~Kalia, A.~Vandenberg-Rodes and D.~Whiteson,
  \emph{{Modeling Smooth Backgrounds and Generic Localized Signals with
  Gaussian Processes}},  \href{https://arxiv.org/abs/1709.05681}{{\ttfamily
  1709.05681}}.

\bibitem{Bevins:2020jqf}
H.T.J.~Bevins, W.J.~Handley, A.~Fialkov, E.~de~Lera~Acedo, L.J.~Greenhill and
  D.C.~Price, \emph{{MAXSMOOTH: rapid maximally smooth function fitting with
  applications in Global 21-cm cosmology}},
  \href{https://doi.org/10.1093/mnras/stab152}{\emph{Mon. Not. Roy. Astron.
  Soc.} {\bfseries 502} (2021) 4405}
  [\href{https://arxiv.org/abs/2007.14970}{{\ttfamily 2007.14970}}].

\bibitem{deLeraAcedo:2022kiu}
E.~de~Lera~Acedo et~al., \emph{{The REACH radiometer for detecting the 21-cm
  hydrogen signal from redshift z\,\ensuremath{\approx}\,7.5\textendash{}28}},
  \href{https://doi.org/10.1038/s41550-022-01709-9}{\emph{Nature Astron.}
  {\bfseries 6} (2022) 984} [\href{https://arxiv.org/abs/2210.07409}{{\ttfamily
  2210.07409}}].

\bibitem{PhysRevD.100.084041}
R.~Buscicchio, E.~Roebber, J.M.~Goldstein and C.J.~Moore, \emph{Label switching
  problem in bayesian analysis for gravitational wave astronomy},
  \href{https://doi.org/10.1103/PhysRevD.100.084041}{\emph{Phys. Rev. D}
  {\bfseries 100} (2019) 084041}.

\bibitem{Bahr:2008pv}
M.~Bahr et~al., \emph{{Herwig++ Physics and Manual}},
  \href{https://doi.org/10.1140/epjc/s10052-008-0798-9}{\emph{Eur. Phys. J. C}
  {\bfseries 58} (2008) 639} [\href{https://arxiv.org/abs/0803.0883}{{\ttfamily
  0803.0883}}].

\bibitem{Bellm:2015jjp}
J.~Bellm et~al., \emph{{Herwig 7.0/Herwig++ 3.0 release note}},
  \href{https://doi.org/10.1140/epjc/s10052-016-4018-8}{\emph{Eur. Phys. J. C}
  {\bfseries 76} (2016) 196}
  [\href{https://arxiv.org/abs/1512.01178}{{\ttfamily 1512.01178}}].

\bibitem{Bierlich:2019rhm}
C.~Bierlich et~al., \emph{{Robust Independent Validation of Experiment and
  Theory: Rivet version 3}},
  \href{https://doi.org/10.21468/SciPostPhys.8.2.026}{\emph{SciPost Phys.}
  {\bfseries 8} (2020) 026} [\href{https://arxiv.org/abs/1912.05451}{{\ttfamily
  1912.05451}}].

\bibitem{Handley:2015vkr}
W.J.~Handley, M.P.~Hobson and A.N.~Lasenby, \emph{{polychord: next-generation
  nested sampling}}, \href{https://doi.org/10.1093/mnras/stv1911}{\emph{Mon.
  Not. Roy. Astron. Soc.} {\bfseries 453} (2015) 4385}
  [\href{https://arxiv.org/abs/1506.00171}{{\ttfamily 1506.00171}}].

\bibitem{Hergt:2021qlh}
L.T.~Hergt, W.J.~Handley, M.P.~Hobson and A.N.~Lasenby, \emph{{Bayesian
  evidence for the tensor-to-scalar ratio $r$ and neutrino masses $m_\nu$:
  Effects of uniform vs logarithmic priors}},
  \href{https://doi.org/10.1103/PhysRevD.103.123511}{\emph{Phys. Rev. D}
  {\bfseries 103} (2021) 123511}
  [\href{https://arxiv.org/abs/2102.11511}{{\ttfamily 2102.11511}}].

\bibitem{Choudalakis:2011qn}
G.~Choudalakis, \emph{{On hypothesis testing, trials factor, hypertests and the
  BumpHunter}},  in \emph{{PHYSTAT 2011}}, 1, 2011
  [\href{https://arxiv.org/abs/1101.0390}{{\ttfamily 1101.0390}}].

\bibitem{Handley:2019tkm}
W.~Handley, \emph{{Curvature tension: evidence for a closed universe}},
  \href{https://doi.org/10.1103/PhysRevD.103.L041301}{\emph{Phys. Rev. D}
  {\bfseries 103} (2021) L041301}
  [\href{https://arxiv.org/abs/1908.09139}{{\ttfamily 1908.09139}}].

\bibitem{Alvarez:2022kfr}
E.~Alvarez, \emph{{Bayesian inference to study a signal with two or more
  decaying particles in a non-resonant background}},
  \href{https://arxiv.org/abs/2210.07358}{{\ttfamily 2210.07358}}.

\bibitem{Kasieczka:2021xcg}
G.~Kasieczka et~al., \emph{{The LHC Olympics 2020 a community challenge for
  anomaly detection in high energy physics}},
  \href{https://doi.org/10.1088/1361-6633/ac36b9}{\emph{Rept. Prog. Phys.}
  {\bfseries 84} (2021) 124201}
  [\href{https://arxiv.org/abs/2101.08320}{{\ttfamily 2101.08320}}].

\bibitem{Yallup:2022yxe}
D.~Yallup, T.~Jan\ss{}en, S.~Schumann and W.~Handley, \emph{{Exploring phase
  space with Nested Sampling}},
  \href{https://doi.org/10.1140/epjc/s10052-022-10632-2}{\emph{Eur. Phys. J. C}
  {\bfseries 82} (2022) 8} [\href{https://arxiv.org/abs/2205.02030}{{\ttfamily
  2205.02030}}].

\bibitem{anesthetic}
W.~Handley, \emph{anesthetic: nested sampling visualisation},
  \href{https://doi.org/10.21105/joss.01414}{\emph{The Journal of Open Source
  Software} {\bfseries 4} (2019) 1414}.

\bibitem{fgivenx}
W.~Handley, \emph{fgivenx: Functional posterior plotter},
  \href{https://doi.org/10.21105/joss.00849}{\emph{The Journal of Open Source
  Software} {\bfseries 3} (2018) }.

\end{thebibliography}\endgroup

\end{document}